\documentclass[10pt,reprint]{iopart}

\usepackage{iopams}  

\usepackage{cite}
\usepackage{times}

\usepackage{epsfig,wrapfig, graphicx,subfigure,multicol}
\usepackage{epstopdf} 

\usepackage{fancyhdr}

\usepackage[margin=2.27cm]{geometry}

\def\Re{{\rm Re\,}}
\def\Im{{\rm Im\,}}

\begin{document}

\title[Reconfigurable reflecting metasurfaces with memory]{Analytical
  and numerical modeling of reconfigurable reflecting metasurfaces with
  capacitive memory}

\author{Abdelghafour Abraray$^{1,2}$, Diogo Nunes$^2$ and Stanislav Maslovski$^{1,2}$}
\address{$^1$ Instituto de Telecomunica\c{c}\~{o}es, Aveiro, Portugal}
\address{$^2$ Dept. of Electronics, Telecommunications and Informatics,
	University of Aveiro, Aveiro, Portugal}

\eads{\mailto{a.abr@av.it.pt}, \mailto{stanislav.maslovski@ua.pt}}
\vspace{10pt}

\begin{abstract}
  In this article, we develop analytical-numerical models for
  reconfigurable reflecting metasurfaces formed by
  chessboard-patterned arrays of metallic patches. These patch arrays
  are loaded with varactor diodes in order to enable surface impedance
  and reflection phase control. Two types of analytical models are
  considered. The first model based on the effective medium approach
  is used to predict the metasurface reflectivity. The second model is
  the Bloch wave dispersion model for the same structure understood as
  a two-dimensional transmission line metamaterial. The latter model
  is used to study ways to suppress parasitic resonances in
  finite-size beamforming metasurfaces. We validate the developed
  analytical models with full-wave numerical simulations. Finally, we
  outline a design of the metasurface control network with capacitive
  memory that may allow for independent programming of individual unit
  cells of the beamforming metasurface.
\end{abstract}

\vspace{2pc}
\noindent{\it Keywords}: metasurface, high impedance surface, surface
impedance,  transmission line metamaterial, Bloch waves


\section{Introduction}

Metasurfaces (MS) are two-dimensional (2D) versions of metamaterials
(MM). In the recent years, they attracted a great deal of attention
due to unprecedented abilities in manipulating the wavefronts of
transmitted and reflected electromagnetic (EM) waves. As compared with
the bulk, three-dimensional (3D) MM, MS typically have negligible
thickness-to-wavelength ratio. Various MS types have been proposed to
transform incident EM radiation in a multitude of ways, with promising
realizations at the microwave, terahertz, and optical
frequencies~\cite{Glybovski}.

In passive MS, the characteristics of the scattered and transmitted
light depend on the geometry, composition and arrangement of the MS
meta-atoms or unit cells, which are fixed at the time of
manufacturing. Thus, the functionality of such MS cannot be changed
once the design is completed. In contrast, with tunable or
programmable MS (PMS), one can realize flexible functionalities by
dynamically controlling the MS interactions with the incident EM waves,
which has a much greater application potential.

Different methods, materials and structures have been proposed to
realize tunable MS at microwave frequencies. For example, tunability
can be achieved by embedding nonlinear elements such as PIN diodes
\cite{Cui2014, Wan2016, Li2017, Li2016, Yang2016}, varactor diodes
\cite{Zhang2018a, Zhang2018b}, semiconductor or micro-mechanical (MEMS)
switches \cite{Ren2019}, etc. While the reconfigurable PMS for
microwave applications are usually controlled electronically, e.g., by
varying the bias voltages on the nonlinear elements, there have been
proposals to use the visible or infrared light for the MS control
\cite{Zhang2018a,lightMM}. At shorter wavelengths, similar functionalities can
be realized by incorporating active or nonlinear material layers into
the MS, such as graphene ribbons \cite{Soleymani2019}, liquid
crystals\cite{Wang2019}, or phase-change materials (e.g., vanadium
dioxide VO$_2$ \cite{Cai2018} and germanium antimony telluride
Ge$_2$Sb$_2$Te$_5$ or GST \cite{Pogrebnyakov2018}).

Nowadays, there is an emerging trend in using Artificial Intelligence
(AI) techniques with the MS designed for high-frequency
applications. The use of AI concepts for beamforming and beam tracking
applications has been proposed before \cite{Hougne2019, Southall1995,
  Zooghby2000, Du2002, Rawat2012}.  However, such solutions
implemented the neural network algorithms solely in software. Because
smart MS can perform practically arbitrary operations on the impinging
wave fronts, they can be employed as trainable hardware components in
artificial neural networks for beamforming
applications~\cite{MM2020,ConfTele2021,MM2021, EExPolytech2021}.

\begin{figure}[htb]
	\centering
	\includegraphics[width=0.6\textwidth]{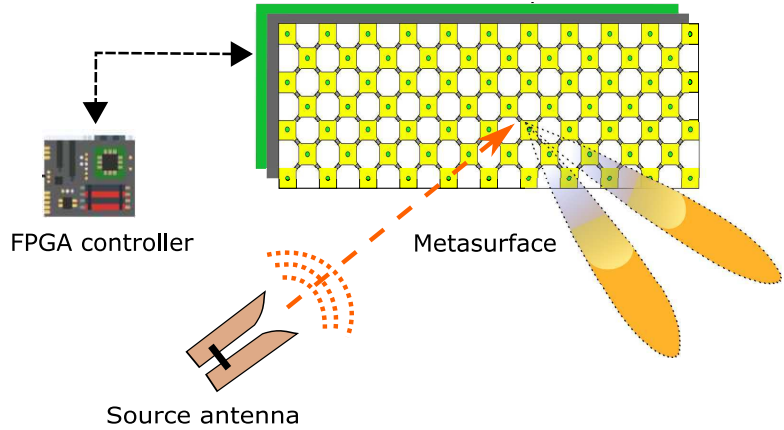}
	\caption{Concept of using reconfigurable reflecting
          metasurface for beamforming applications. }
	\label{fig: MS_topology_inksc}
\end{figure}

MS targeted for applications at the microwaves are traditionally
formed by arrays of electrically small elements, which are
metal-dielectric structures with the geometry and composition selected
carefully in order to generate the needed EM response.
In particular, reflecting PMS of different
designs have been actively discussed in the literature. Such PMS can
be used as controlled reflectors in beamforming antennas.
Figure~\ref{fig: MS_topology_inksc} shows the main concept of using
reconfigurable reflecting MS in wireless communications.

Due to a high interest in the reconfigurable MS for telecommunication
applications, it is important to search for accurate analytical and
numerical models of such structures. Because here we will deal with
the models of smart reflecting MS, our goal will be to provide a set
of analytical and numerical tools suitable for a) modeling the
reflection properties of the periodic MS with embedded controlling
elements within the framework of the local surface impedance
approximation; and b) identifying the parasitic surface mode
resonances in such structures and proposing the methods of their
suppression. It is hard to underestimate the importance of the surface
mode management in the beamforming MS. In fact, an unprecedented level
of control over the MS reflection patterns can be achieved with proper
``surface mode engineering'' techniques that have been developed only
relatively recently~\cite{DiRenzo2020,perfectrefl,anomalous,RaadiPRL,metagratings,Metagratings}.

In this article, we consider a high impedance surface-based (HIS-based) design that
utilizes an array of metallic patches with vias placed above a metal
plane, similar to the Sievenpiper's mushrooms~\cite{Sievenpiper1999,Sievenpiper2003}. However, we arrange the
patches in a chessboard-like structure with the nonlinear capacitive
loads (varactors) connected to the neighboring patches at the patch
corners. This change in the unit cell geometry allows us to increase
the resonant frequency of the HIS. In our structure, the resonant
frequency also appears to be stable versus the incidence angle,
independently of the incident wave polarization.

An analytical model for such reflecting MS can be developed
using a quasistatic model for the uniaxial wire media~\cite{MOTL,WM,WM_PRB} and the
additional boundary conditions at the patch-to-via
interfaces~\cite{ABC}. However, such models assume that the
MS is regular (periodic) in its own plane. In order to
account for irregularities (e.g., at the MS edges) we also
develop an alternative analytical model based on the dispersion
equation for Bloch waves in a two-dimensional (2D) transmission line (TL)
metamaterial, with which we study the surface wave excitation and
suppression in such MS.

We validate the developed analytical models with numerical simulations
in CST Studio Suite and Agilent ADS. In this article we also consider a
possible realization of the controlling network that includes
capacitors that are used as analog memory to program the MS
configuration. Our goal is to keep this network as simple as possible,
while realizing independent control of every element of the MS.

\section{Analytical model of a periodic varactor-loaded mushroom-type
  MS}
\label{imped_model}

The so-called mushroom-type HIS was first proposed in 1999 by
Sievenpiper \cite{Sievenpiper1999}. Since then analogous designs have
been employed in a great number of works (see, for instance,
~\cite{Sievenpiper1999, Sievenpiper2003, Yakovlev2009,
  book_tretyakov}). Mushroom structures can be realized as arrays of metal
patches of different shapes (square, hexagonal, Jerusalem crosses,
etc.) placed above a grounded dielectric slab (typically, a printed
circuit board (PCB) substrate). In such HIS, the patches are connected
to the ground by metallic vias passing through the dielectric
substrate. The vias operate as an effective wire medium (WM, \cite{WM}) layer
inserted between the patches and the ground. It can be shown that the
WM layer stabilizes the resonant frequency and impedance of the HIS for
the waves of transverse magnetic (TM) polarization. The presence
of vias is also convenient for realizing loading and controlling
networks.

\begin{figure}[t]
	\centering
	\includegraphics[height=0.3\textwidth]{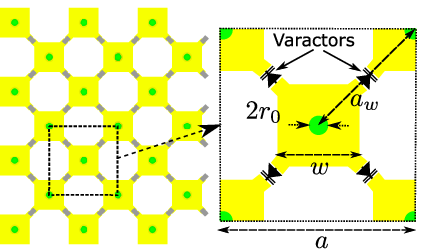}\hspace{0.5cm}
        \includegraphics[height=0.33\textwidth]{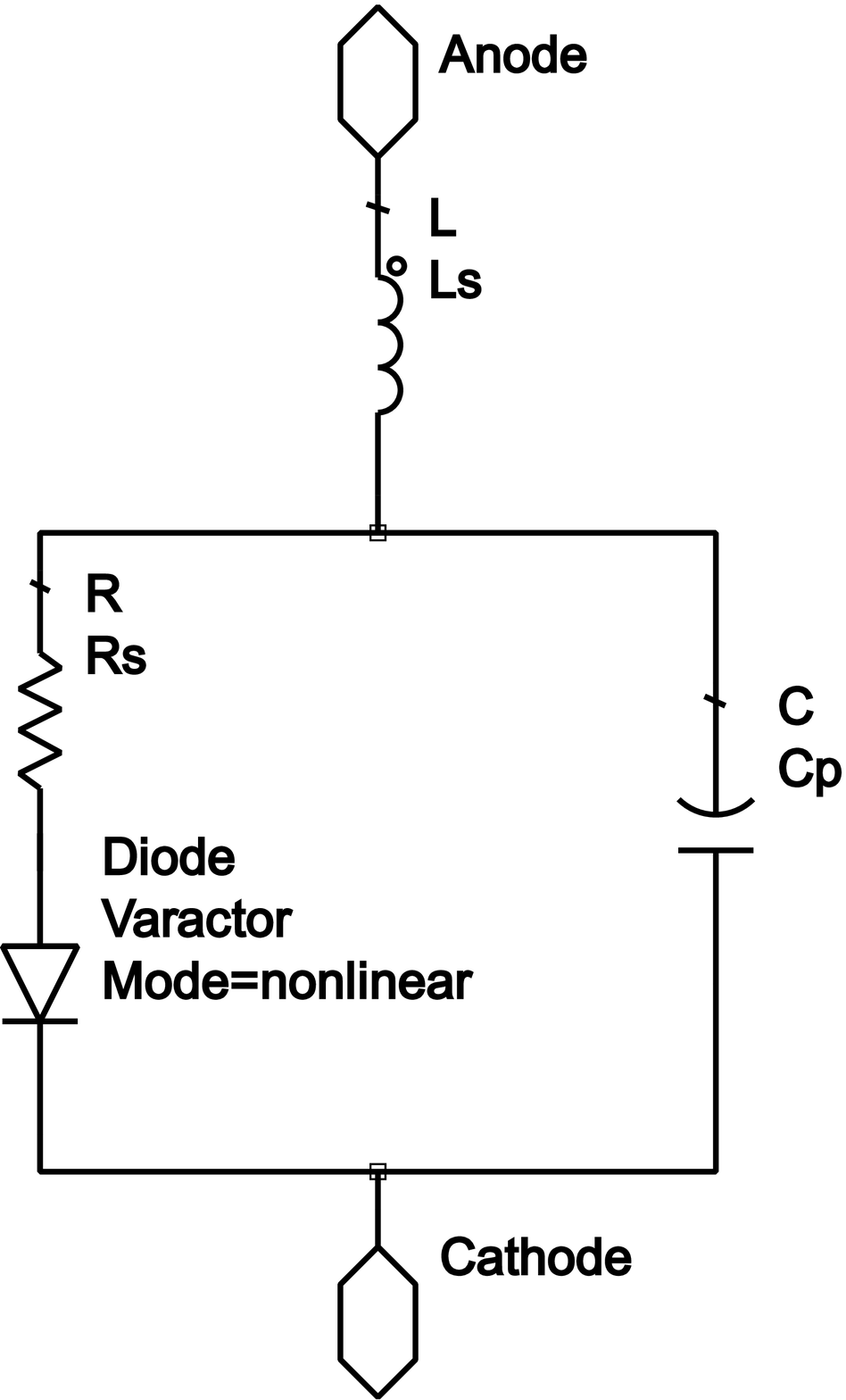}\\
        \hspace{2cm}(a)\hspace{6.2cm}(b)
	\caption{Panel (a): Top view of the chessboard mushroom-type
		artificial impedance surface with varactor diodes. The
		green circles mark the vias connecting the square metallic
		patches to the controlling lines located below the ground
		plane (not shown). The dashed square delimits a
		unit cell of the structure, which is also shown in the
                inset. Panel (b): SPICE model of the varactor diode.}
	\label{fig:spice_model}
\end{figure}

In what follows, we analyze a reflecting MS based on a
capacitively-loaded mushroom-type HIS whose top view is depicted in
Figure~\ref{fig:spice_model}. The MS structure shown in this figure is
based on a chessboard-like array of square metallic patches with
nonlinear capacitive loads (varactors) connected between the corners
of the neighboring patches. The array period is $a$.  At the middle of
each unit cell, there is a via with radius $r_0$ that passes through a
hole in the ground plane. Below the ground plane, this via is
connected to a dc bias line (the $x$-coordinate control line) with
negative polarity. Analogously, after passing trough holes in the
ground plane, the four vias at the corners of the unit cell are
connected to the bias lines with positive polarity (the $y$-coordinate
control lines).

In a practical design, the control lines can be
formed on a pair of separate metalization layers in a multilayer PCB,
or on a separate control PCB located below the ground plane of the HIS
structure. In such configuration, the controlling lines do not
interfere with the high-frequency fields at the front side of the
MS. In order to even better isolate the controlling network from the
induced high-frequency currents, one may also place filtering capacitors
between the control lines and the HIS ground plane.

From the above description, it is apparent that such a structure can
operate as a tunable MS, in which varying voltages applied to the
control lines may change the resonant frequencies of the MS unit
cells. With a simple modification of the controlling network, one can
also make every unit cell of such MS independently programmable by
introducing memory capacitors, which will be explained in detail in
Section~\ref{control_circuit}.

In the rest of this section, we derive an analytical model for the effective
surface impedance and the reflection coefficient for such MS in a
linear response regime, i.e., when the amplitude of the high-frequency
voltage across the varactors is much smaller than the respective
reverse bias voltage. This situation is easily attainable in an MS
operating as a tunable reflector in a receiving antenna. This assumption
can also hold in low-power transmitting systems.

Under these assumptions, when the period of the MS, $a$, is small as
compared to the wavelength, $\lambda$, we may treat the MS as a
locally homogeneous periodic structure with some effective surface
impedance. This impedance has to be determined from the geometry of
the unit cells and the complex impedance of the varactors. By itself,
a dense array of disconnected patches acts as an effective capacitive
impedance.  Due to the induced high frequency currents flowing along
the patch edges there exists also an inductive component in the patch
grid impedance.

The effective grid capacitance $C_g$ (normalized to $\epsilon_0$) and
the inductance $L_g$ (normalized to $\mu_0$) for a simple periodic grid
of square patches can be expressed as
follows (see, e.g., \cite{book_tretyakov}):
\begin{equation}
\label{eq:3_grid_capacitance_inductance}
{C_g} =
\frac{(\varepsilon_r+1)w\log{\csc{\frac{\pi(a_e-w)}{2 a_e}}}}{\pi},\qquad
{L_g} = 
\frac{w\log{\csc{\frac{\ensuremath{\pi}  \left( a_e-w\right) }{2 a_e}}}}{2 \pi },
\end{equation}
where $a_e$ is the grid period. We have found that these formulas also
work well for a chessboard-like grid of patches, if one finds a
suitable effective value for $a_e$. Namely, with a series of numerical
tests we have found that the value $a_e = a/2 \pm \delta$, where
$\delta \ll a_e$ is a small correction, works quite well for the
chessboard-like MS considered in this article. The value of $\delta$
can be determined from numerical simulations for the complex
reflection coefficient of the MS, by matching the resonant frequency
predicted by the analytical model to the one obtained from the
simulations. After such a correction is applied, the results of the
analytical model are in excellent agreement with the simulation
results in a wide frequency range and for arbitrary incidence angles
(see Section \ref{num_results}).

Under this approximation, the complex admittance $Y_g$ of the
capacitively loaded chessboard-like patch grid (normalized to
$\eta_0^{-1} = \sqrt{\varepsilon_0/\mu_0}$, where $\eta_0$ is the
free space impedance), can be expressed as
\begin{equation}
\label{eq:3_eq_admittance_Yg}
Y_g = \frac{1}{j k_0 L_g+\frac{1}{jk_0 C_g+Y_d}},
\end{equation}
where we have taken into account the complex varactor admittance,
$Y_d$ (normalized to $\eta_0^{-1}$). Because varactors connect
the neighboring patches, this admittance is effectively in parallel
with the grid capacitance $C_g$. Here and in what follows,
$k_0 = \omega\sqrt{\varepsilon_0\mu_0}$ is the free space
wavenumber. The admittance $Y_d$ can be obtained from the equivalent
SPICE model of the varactor diode [Figure~\ref{fig:spice_model}(b)]:
\begin{equation}
\label{eq:3_eq_admittance_Yd}  
{Y_d}= \frac{1}{j k_0 L_s +\frac{1}{j k_0 C_p +
		\frac{1}{R_s+\frac{1}{j k_0 C_j}}}},
\end{equation}
where $C_j$ is the normalized junction capacitance of the varactor
diode, $C_p$ is the normalized package capacitance, $L_s$ is the
normalized serial inductance, and $R_s$ is the effective serial resistance
(normalized to $\eta_0$).

\begin{figure}[!t]
	\centering
	\includegraphics[height=3.5cm]{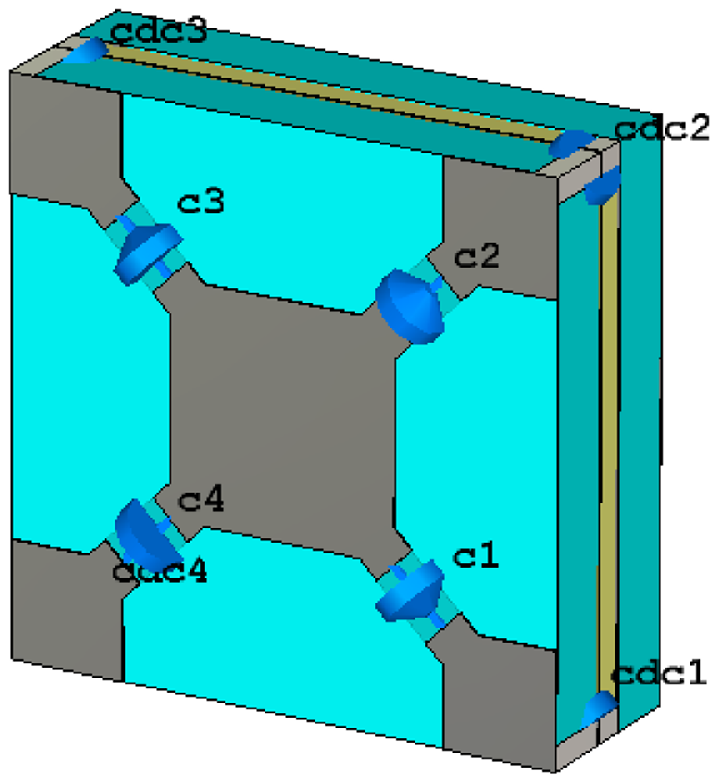}\hspace{0.7cm}
	\includegraphics[height=2.7cm]{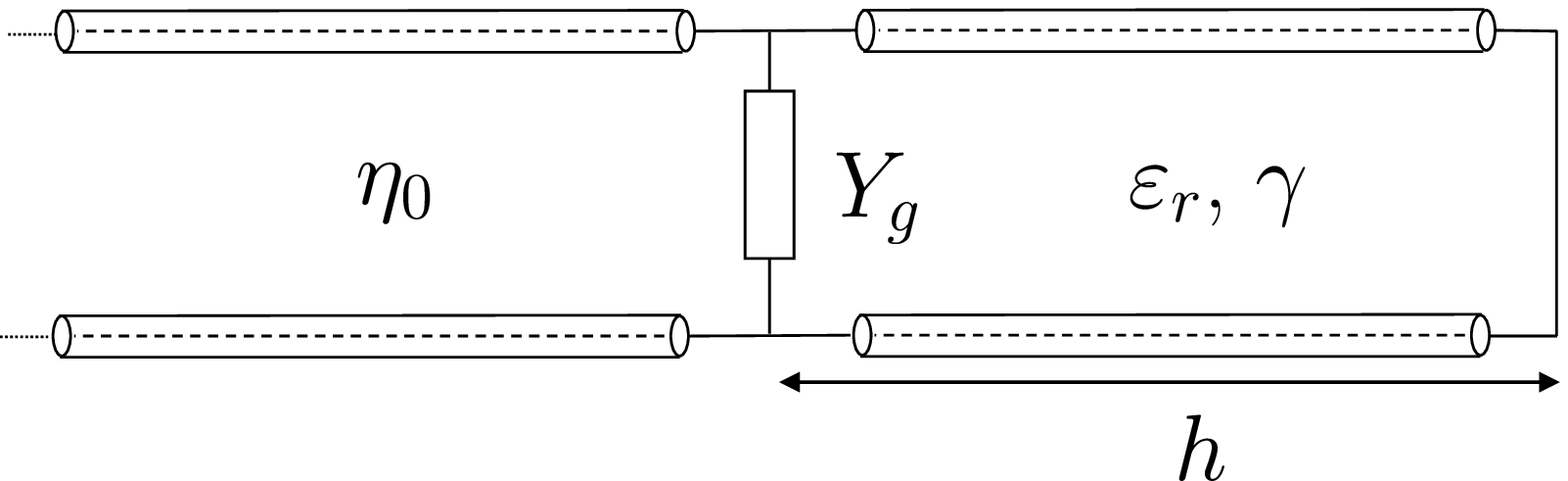}\\
	\hspace{-3cm}(a)\hspace{6.5cm}(b)\\
	\caption{Panel (a): The unit cell of the PMS structure as
		modeled in SIMULIA CST Studio Suite. Panel (b):
		TL model for the metal-backed HIS. Here, $h$ is the substrate thickness,
		$\varepsilon_r$ is the substrate relative permittivity,
		$\gamma = \gamma^{TM,TEM,TE}$ is the propagation constant of
		a given mode, $\eta_0$ is the free space impedance, and
		$Y_g$ is the effective admittance of the patch grid. }
	\label{fig:transmission-line_model}
\end{figure}

Let us first consider the TM incidence. In this case, the normalized
surface admittance $Y_s^{TM}$ of the capacitively loaded
chessboard-like MS can be obtained with an equivalent TL model shown in
Figure~\ref{fig:transmission-line_model}(b). From this figure it is
evident that the admittance $Y_s^{TM}$ is a parallel connection of the
surface admittance of the patch grid $Y_g$ and the input admittance of
the WM slab $Y_{wm}$:
\begin{equation}
\label{eq:3_total_admittance}
{Y^{TM}_s}  = {Y_{wm}+Y_g}.
\end{equation}

Unlike the typical case of a dielectric slab under the TM incidence,
in our situation there are {\em two} modes propagating inside the
uniaxial WM slab formed by the dielectric substrate and the vias: the
TM mode and the TEM (transverse electro-magnetic) mode. The complex propagation factors for these
waves can be expressed as follows~\cite{WM}.  For the TM
wave:
\begin{equation}
\label{eq:3_gamma_TM}
\gamma^{TM}=\sqrt{{k_t^{2}}+{k_p^{2}}-\varepsilon_r\, {k_0^{2}}},
\end{equation}
and for the TEM wave:
\begin{equation}
\label{eq:3_gamma_TEM}
\gamma^{TEM} = j{k_0} \sqrt{\varepsilon_r},
\end{equation}
where $k_t = k_0\sin\theta$ is the transverse wavenumber (with $\theta$
being the angle of incidence), $k_p$ is the WM plasma wavenumber,
$\varepsilon_r$ is the relative permittivity of the substrate. The plasma
wavenumber ${k_p}$ for the uniaxial WM can be approximated as~\cite{WM_PRB}
\begin{equation}
\label{eq:3_plasma_wavenumber}
{k^2_p} = 
\frac{2 \ensuremath{\pi} }{a_w^2 \log{\frac{{a_w^{2}}}{4 \left( a_w-{r_0}\right) \, {r_0}}}},
\end{equation}
where $a_w = a/\sqrt{2}$ is the shortest distance between the vias (the WM period).

Next, the normalized input admittance of a uniaxial WM slab, $Y_{wm}$,
with the wires terminated by patches at one side of the slab and
connected to the ground at the other side of the slab can be derived
by considering the above-mentioned modes and the complete set of
boundary conditions (which includes additional boundary conditions,
ABCs) at the two sides of the slab, in a manner similar to what was
done in~\cite{ABC}. After some algebra, the admittance
$Y_{wm}$, as is seen from the side of the WM slab adjacent to the
patch grid, can be expressed as
\begin{equation}
\label{eq:3_admittance_Ywm}
{Y_{wm}}=
\frac{j\varepsilon_rk_0\left(\kappa\cosh h\gamma^{TM} + \cosh
	h\gamma^{TEM}\right)}{\kappa \gamma^{TM} \sinh
	h\gamma^{TM} +\gamma^{TEM} \sinh h\gamma^{TEM}},
\end{equation}
where the parameter $\kappa$ can be expressed as 
\begin{equation}
\label{eq:3_kappa}
\kappa=\frac{k_t^{2}}{k_p^{2}}\frac{\cosh{h\gamma^{TEM}} + \alpha
	\gamma^{TEM}\sinh{h\gamma^{TEM}}}{
	\alpha\gamma^{TM}\sinh{h\gamma^{TM}} + \cosh{h \gamma^{TM}}}.
\end{equation}
In these expressions, $h$ is the substrate thickness, and $\alpha$ is the ABC parameter~\cite{ABC}. In this model, we are taking into
account the ABC at the connection of vias to the patches, and thus
$\alpha$ can be expressed as follows
\begin{equation} 
\label{eq:3_aplpha_parameter}
\alpha = 
\frac{\left( \varepsilon_r+1\right) w \log{\frac{{a_w^2}}{4 \left(a_w-r_0\right)r_0} } }{2 \varepsilon_r \log{ \sec{ \frac{\pi  \left(a_e-w\right) }{2 a_e} } }}.
\end{equation}

On the other hand, the same admittance can be approximated by
considering only the TEM waves in the WM slab and neglecting the TM
waves (which are evanescent when $\sqrt{\varepsilon_r}k_0 <
k_p$). Under this approximation, the normalized input admittance $Y_s^{TEM}$ reads
\begin{equation}
\label{eq:3_total_admittance_TEM}
{Y^{TEM}_s} = Y_g - j\sqrt{\varepsilon_r}\cot{\left( \sqrt{\varepsilon_r} k_0h\right)}.
\end{equation}

Let us now consider the transverse electric (TE) incidence. In this case, the only modes
existing in the WM slab are the TE waves with the complex propagation
factor
\begin{equation}
\label{eq:3_gamma_TE}
\gamma^{TE}=\sqrt{{k_t^{2}}-\varepsilon_r\, {k_0^{2}}},
\end{equation}
because under the TE incidence the thin vias practically do not
interact with the incident electric field (because it is perpendicular
to the vias). Thus, for the TE incidence, the normalized input
admittance can be written as
\begin{equation}
\label{eq:3_total_admittance_TE}
{Y^{TE}_s} = Y_g + \frac{\gamma_{TE}\coth{\left( \gamma^{TE} h\right) }}{jk_0}.
\end{equation}

Finally, the electric field reflection coefficient for the cases of
the TM and TE incidence can be defined by using the full model as
\begin{equation}
\label{eq:3_reflecton_coefficient}
R_s^{TM} = 
\frac{1-Y_s^{TM}\cos\theta}{1+Y_s^{TM}\cos\theta}, \qquad
R_s^{TE} = 
\frac{1-Y_s^{TE}/\cos\theta}{1+Y_s^{TE}/\cos\theta}.
\end{equation}
Alternatively, when using the simplified model for the TM case,
\begin{equation}
R_s^{TM} = 
\frac{1-Y_s^{TEM}\cos\theta}{1+Y_s^{TEM}\cos\theta}.
\end{equation}
The simplified model can be used for validation of the complete model at the frequencies where the TM mode in the WM has no effect. 

\section{Numerical modeling of the PMS
  equipped with a control board}
\label{num_results}

In order to validate the analytical model, we have simulated the
considered PMS structure in the commercial EM full-wave
simulation software SIMULIA CST Studio Suite. In addition, in the
numerical simulations we have considered a more realistic structure
that included a control board that is placed at the back of the HIS.

As it was mentioned before, the considered PMS can be formed by a pair
of PCBs: the frontside board that realizes the mushroom-type HIS, and
the backside board with the $x$ and $y$ control lines. We have built
one period of such a structure (the unit cell) in the 3D modeler of
the SIMULIA CST Studio Suite. Realistic material models for the PCB
substrates (Isola IS680 with the thickness $h = 1.52$~mm and the
relative permittivity $\varepsilon_r = 3.38$, $\tan\delta = 0.002$)
and metalization layers (annealed copper with thickness $t = 35$~um,
$\sigma = 5.8\times10^7$ S/m) have been used. A view of the unit cell
of this structure as realized in SIMULIA CST Studio Suite is shown in
Figure~\ref{fig:transmission-line_model}(a). The two PCBs are
separated by a gap of $0.5$~mm filled with epoxy resin
($\varepsilon_r = 4$). The period of the MS was $a = 6.8$ mm and the
patch size was $w = 3$~mm. The varactor diodes were modeled with a
lumped $LCR$ equivalent circuit resulting from the SPICE model of the
varactor diode [Figure~\ref{fig:spice_model}(b)], whose capacitance
was changed depending on the studied case, which corresponded to
changing the varactor diode control voltage.  The unit cell model also
included the filter capacitors connected at the points where the vias
pass through the ground plane of the front-side PCB.

The unit cell was placed under the Floquet-periodic boundary
conditions and the incident plane wave propagating at an angle
with respect to the $z$-axis (normal to the MS) was modeled by using a
Floquet port. The Floquet modes that simulated the
incident waves consisted of two orthogonally polarized plane
waves. The reflection amplitude and phase for these incident waves
were recorded as functions of frequency and the incidence angle.

\begin{figure}[h]
	\centering
	\includegraphics[width=0.6\columnwidth]{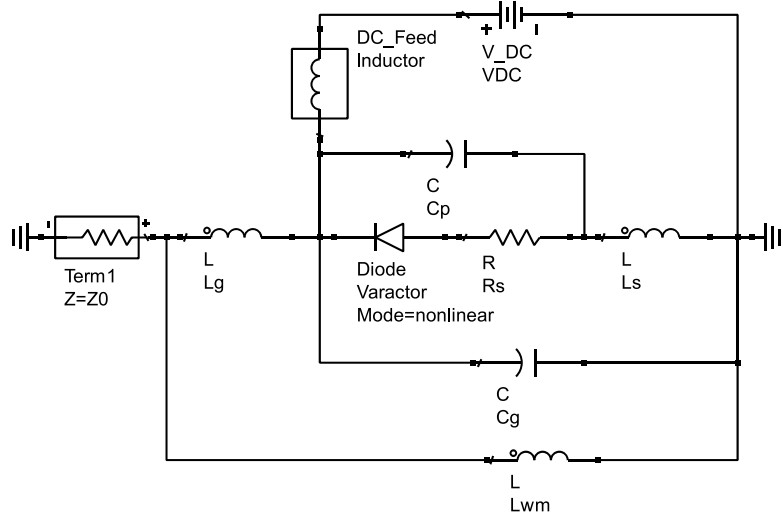}
	\caption{Equivalent circuit model of the varactor-loaded unit cell
          as realized in Agilent ADS.}
	\label{fig:MS_ads_model}
\end{figure}

The MS unit cell model was also simulated using Agilent ADS. In ADS, the MS unit cell can be
modeled with an equivalent circuit dictated by the structure of
Eqs.~(\ref{eq:3_eq_admittance_Yg}), (\ref{eq:3_eq_admittance_Yd}), and
(\ref{eq:3_total_admittance}). The ADS circuit was formed by an
effective inductance $L_{wm} = \eta_0/(\omega_0\,\Im Y_{wm}|_{\omega_0})$
modeling the WM layer (here, $\omega_0 = 2 \pi f_0$, with $f_0$ being
the desired central frequency) connected in parallel to a circuit
modeling the loaded patch grid that comprised a lumped capacitor
$C = \varepsilon_0 C_g$, an inductor $L = \mu_0 L_g$,
[Eq.~(\ref{eq:3_grid_capacitance_inductance})], and a varactor diode
represented by its SPICE model (Figure~\ref{fig:MS_ads_model}).

\begin{figure}
	\centering
	\subfigure[]{\label{fig:phase_varactor}\includegraphics[width=0.45\textwidth]{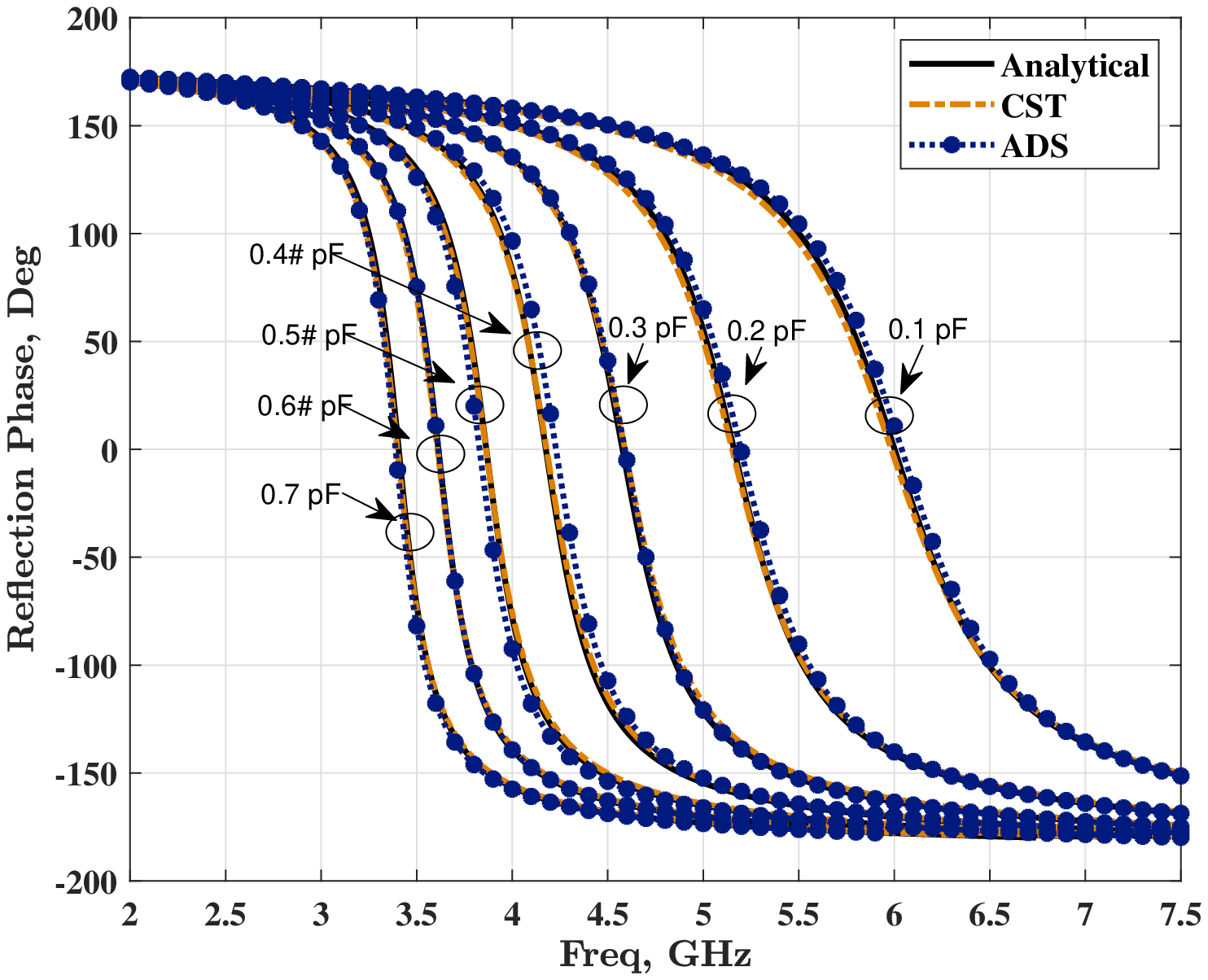}}
	\subfigure[]{\label{fig:magnitude_varactor}\includegraphics[width=0.45\textwidth]{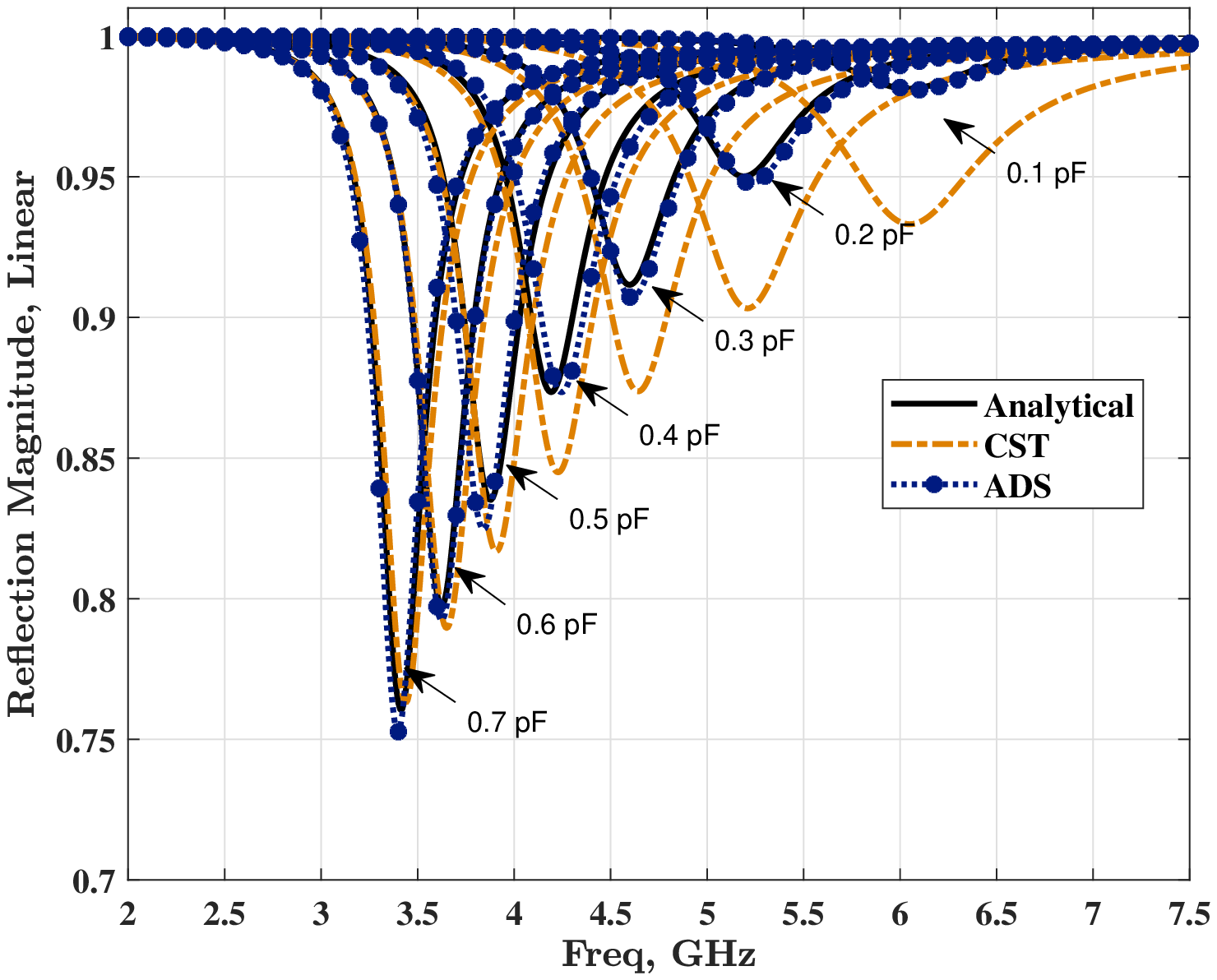}}
	\caption{Complex reflection coefficient versus frequency for
          different values of the junction capacitance corresponding
          to the varactor diode MAVR-000120-1141. Panel (a):
          Reflection phase. Panel (b): Reflection magnitude. The
          varactor diode has the serial inductance $L_s = 0.2$~nH,
          serial resistance $R_s = 0.88$~Ohm, and package capacitance
          $C_p = 0.14$~pF. Under the control voltage variation from 0
          to $V_b = 20$~V (the breakdown voltage) the junction
          capacitance $C_j$ varies from about $1$ to $0.1$~pF. The
          curves are obtained with the correction parameter
          $\delta = 0.3$~mm. The equivalent parameters of the diode
          were confirmed by vector-network analyzer (VNA)
          measurements.}
	\label{fig:s11_normal}
\end{figure}

The phase of the reflection coefficient under normal incidence
obtained from the analytical model and from the CST and ADS
simulations is shown in Figure~\ref{fig:s11_normal} for different
capacitor values. By decreasing the capacitance value from $0.7$ to
$0.1$~pF, the resonant frequency can be tuned over a range from about
$3.4$ to $6$~GHz.  For each value of the capacitance, the reflection
phase varies with the range of about 360$^\circ$ and crosses through
zero at the resonant frequency of the MS, where it behaves as an
artificial magnetic conductor, because close to the resonant frequency
the input impedance of the MS is very high.  In order to achieve efficient
control of the reflected wavefront, the reflection phase
should ideally cover the full 360$^\circ$ range.

\begin{figure}
	\centering
	\subfigure[]{\label{fig:TE2}\includegraphics[width=0.45\textwidth]{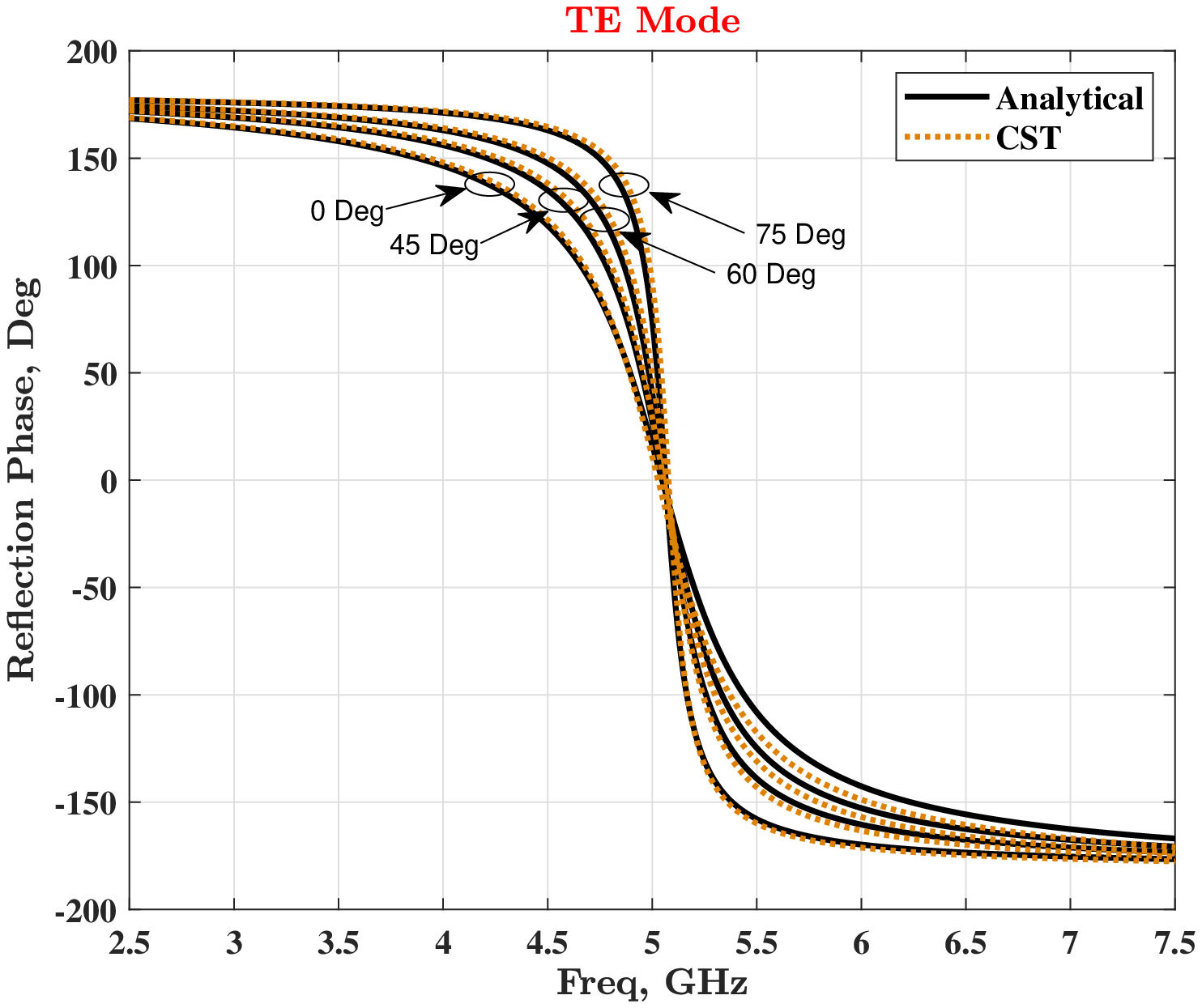}}
	\subfigure[]{\label{fig:TM2}\includegraphics[width=0.45\textwidth]{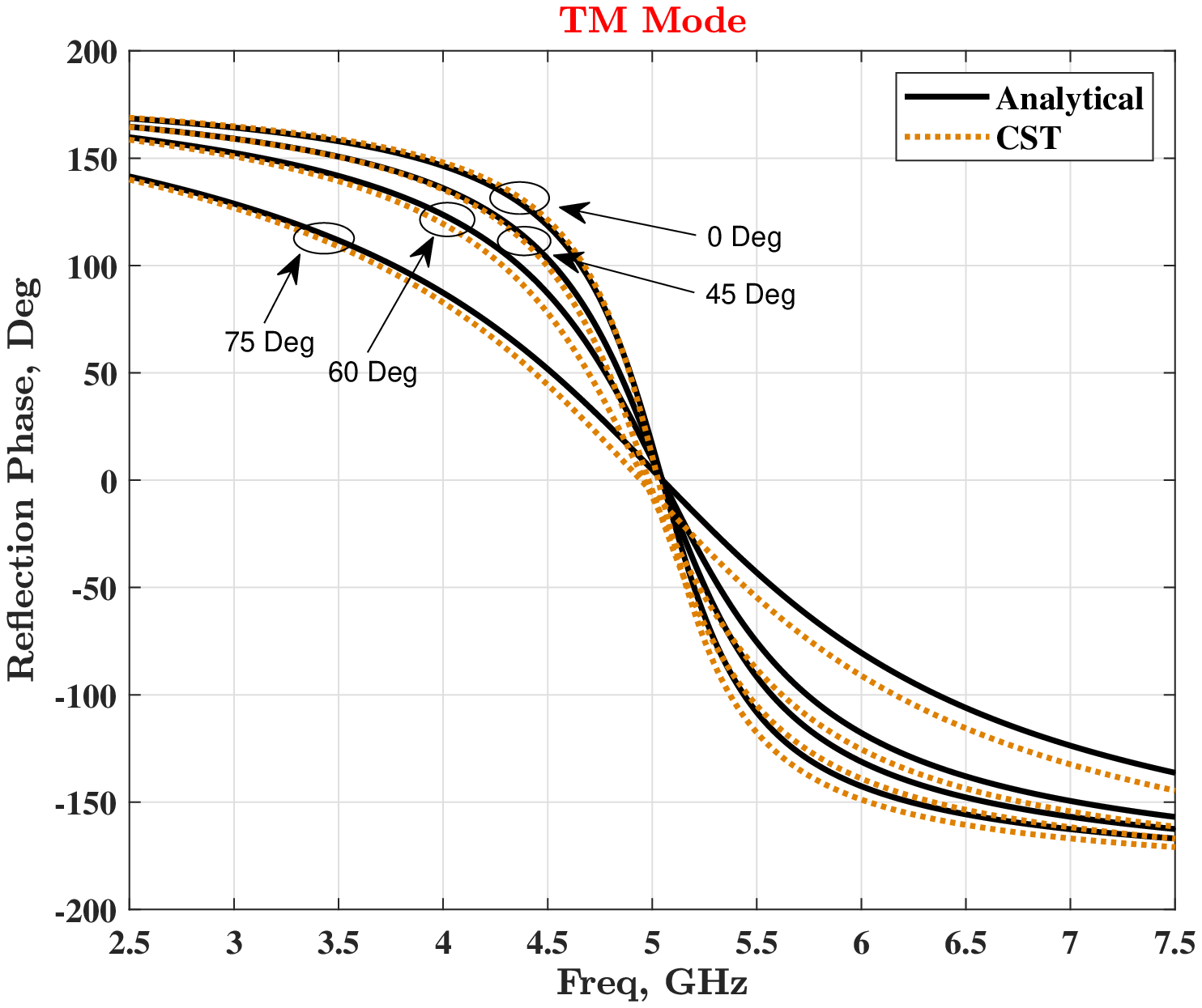}}
	\caption{Reflection phase versus frequency for different
          values of the incidence angle. Panel (a): TE
          incidence. Panel (b): TM incidence.}
	\label{fig:s11_oblique}
\end{figure}

Dependence of the reflection phase on frequency for the same MS
structure has been also studied under oblique incidence. The
corresponding results for the case when the HIS resonates at about 5
GHz are presented in Figure~\ref{fig:s11_oblique}, from which one can
see that although the behavior is different for the TE and TM
polarizations, the reflection phase curves for all incidence angles
cross zero at approximately the same frequency. The resonance
bandwidth decreases for the TE polarization and increases for the TM
polarization as the angle of incidence grows, as expected. Thus we may
conclude that the studied structure has robust performance with
respect to the incident wave polarization and the angle of
incidence. In all studied cases, the results of the analytical model
agreed very well with the numerical simulation results. The numerical
simulations also confirmed that the control board placed behind the
HIS has no effect on the performance of the HIS at the microwave
frequencies.

\section{Nonuniformities and parasitic Bloch waves on finite
  mushroom-type HIS}
\label{Bloch_model}

The analytical model developed in the previous section is valid only
for a uniform MS with equal unit cells distributed periodically over a
very large (theoretically, infinite) area. However, because the
beamforming operation requires varying the local surface impedance
over the MS in a complex way, the beamforming PMS generally cannot be
uniform from the EM point of view. Realistic PMS also
have finite dimensions. Therefore, complementary models or extended
models are needed to describe such generally nonuniform PMS.

There have been successful attempts on modeling MS with periodic
geometries and with nonuniform distributions of surface impedance
among the unit cells. In general, treatment of such structures
requires considering coupling between the Floquet harmonics of the
field scattered by the MS. Beamforming techniques based on such
approaches allow for a full control over the radiation pattern of a
reflecting
MS~\cite{DiRenzo2020,perfectrefl,anomalous,RaadiPRL,metagratings,Metagratings}.
Because the higher-order Floquet harmonics are similar to resonating
surface modes supported by periodic structures, one may say that
beamforming in such structures is realized via proper surface mode
engineering. In theory, patterns with arbitrary main beam directions
and low sidelobe levels can be achieved with such techniques.

However, and especially in finite-size PMS that comprise many complex
unit cells, the surface mode engineering problem may become very
complex, with many influencing factors that might be overlooked by the
available analytical-numerical models. For example, the surface
impedance-based models that have a very good accuracy when dealing
with the plane wave incidence and reflection, may become rather
inaccurate when dealing with non-uniformly excited realistic
MS. Indeed, realistic HIS include several dielectric and metalization
layers. Thus such complex structures may support additional modes that
propagate in these layers. The finite size of the unit cells also
makes the local surface impedance models inaccurate due to the spatial
dispersion effects that are prominent when the wavelength becomes
comparable to the unit cell size.

In this section we are going to investigate the surface
modes that exist in the finite-size chessboard-like mushroom
structures. This study was motivated by unwanted phenomena that we
observed when we numerically simulated finite MS samples with the unit
cell shown in Figure~\ref{fig:transmission-line_model}(a). Namely, in
those simulations we observed excitation of parasitic modes produced
by irregularities at the edges. An efficient approach to tackle such effects in the
finite-size structures can be devised by using the Bloch
wave dispersion theory for the 2D TL-based metamaterials
\cite{Maslovski2018}.

Let us consider a finite MS sample formed by equal unit cells
terminated at the MS edges with some impedances that we need to
determine. In order to suppress the parasitic resonating modes, we may
use the fact that when the termination impedance equals the
characteristic impedance of the unwanted mode, the reflections at the
MS edges are suppressed and the parasitic resonances disappear.

\begin{figure}
	\centering
	\includegraphics[width=0.32\textwidth]{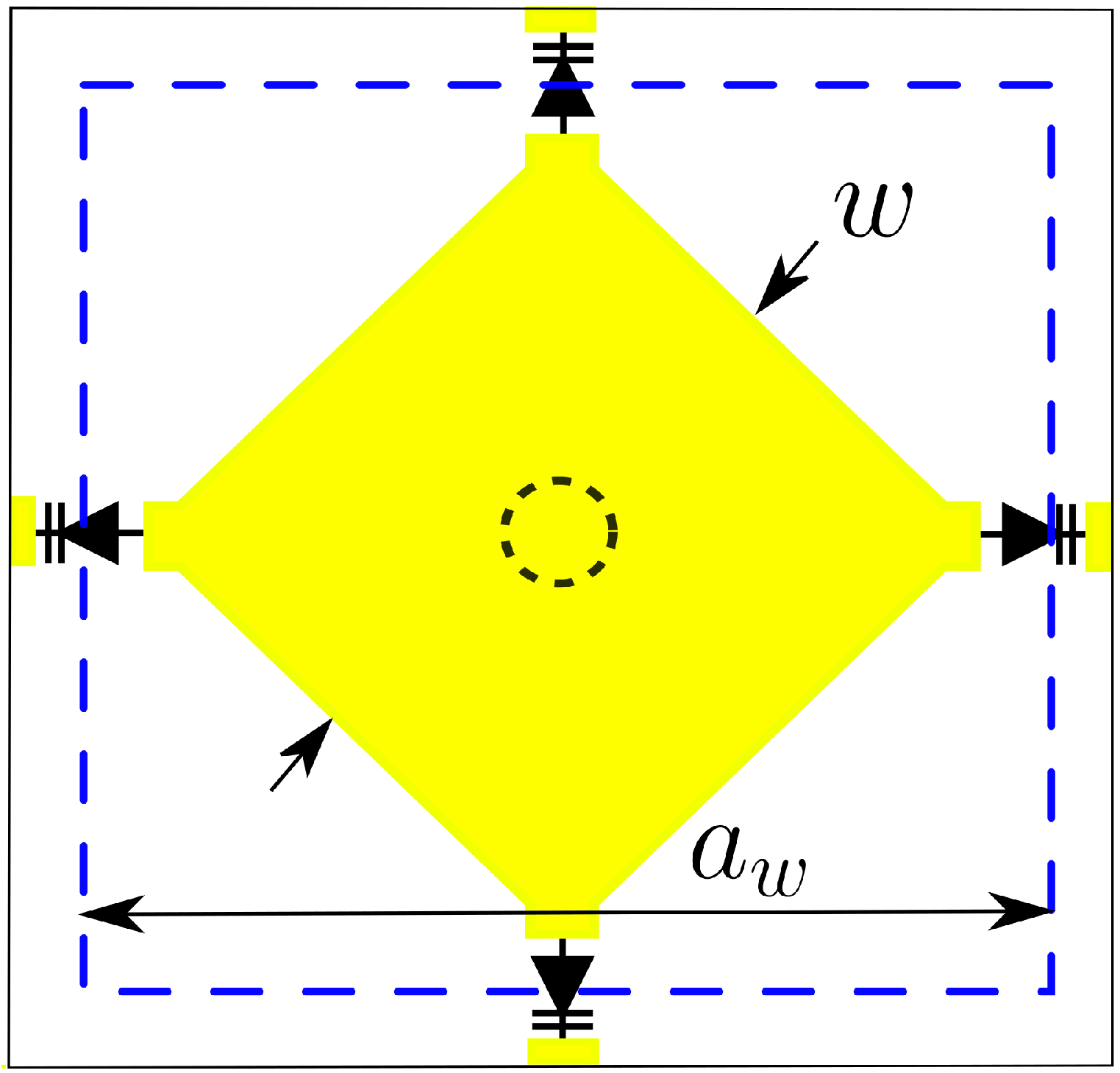}
	\includegraphics[width=0.3\textwidth]{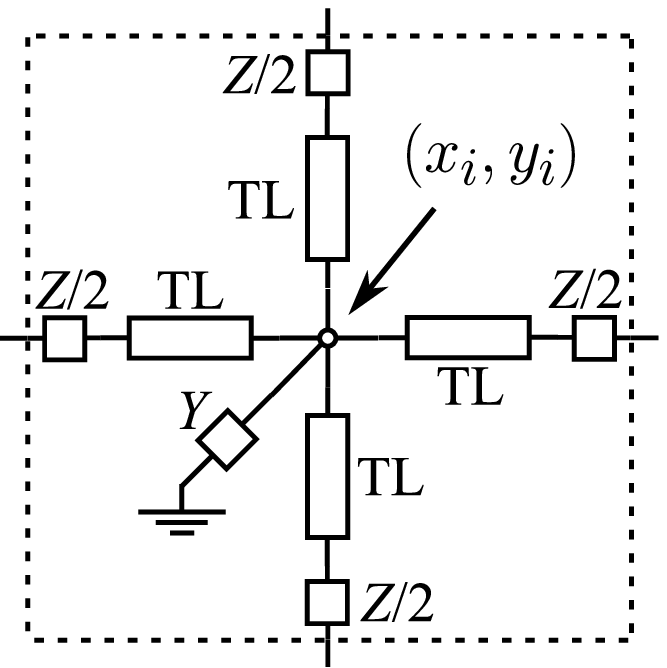}\\
	(a)\hspace{4.5cm}(b)\\
	\caption{Panel (a): Reduced 45$^\circ$-rotated unit cell of the
		chessboard structure (inside the dashed square). Panel (b): Schematic of the
		corresponding TL-based 2D metamaterial.}
	\label{fig:TL_circ}
\end{figure}

To find the dispersion and the impedance of the MS-supported parasitic
modes understood as the Bloch waves in a 2D-periodic TL-based
metamaterial we consider the equivalent circuit of the reduced,
45$^\circ$-rotated unit cell shown in Figure~\ref{fig:TL_circ}. In the
equivalent circuit of this cell, the admittance $Y$ is formed by a
parallel connection of the via's inductance $L_{via}$ to the
ground~\cite{MOTL},
\begin{equation}
\label{eq:3_Lvias}
L_{via} = \frac{\mu_0 h \log\frac{a_w^2}{4r_0(a_w-r_0)}}{2 \pi},
\end{equation}
and the effective patch capacitance to the ground, $C_{pat}$, given by
\begin{equation}
\label{eq:3_Cpatch}
C_{pat} = \frac{\varepsilon_0\varepsilon_rw^2}{h}+C_0,
\end{equation}
where the additional capacitance, $C_0$, is the
self-capacitance of a patch in the periodic array of patches~\cite{WM_PRB}:
\begin{equation}
\label{C_0}
C_0 = \frac{\varepsilon_0\pi w(\varepsilon_r+1)}{\log\sec \frac{\pi(a_e-w)}{2 a_e}}.
\end{equation}
From these results, the shunt admittance $Y$ that
connects every patch to the common ground reads
\begin{equation}
\label{admittance_Y}
Y = \frac{1}{j\omega L_{via}} + j\omega C_{pat}.
\end{equation}
The serial impedance $Z$ represents the small-signal complex
impedance of a varactor diode that can be obtained from the SPICE
model of the diode: $Z = \eta_0 Y_{d}^{-1}$, where $Y_{d}$ is given by
Eq.~(\ref{eq:3_eq_admittance_Yd}).

Due to the 45$^\circ$ rotation of the unit cell, the surface waves
that impinge on the MS edges at the right angle have
$|k_{x'}| = |k_{y'}|$ in the rotated coordinate system $(x',y')$ of
the reduced unit cell shown in Figure~\ref{fig:TL_circ}. Taking this fact into
account, the resulting dispersion equation can be written as~\cite{Maslovski2018}:
\begin{eqnarray}
\label{dispersion_equation_kx}
  k_{x'} = \pm a_w^{-1}\cos^{-1}\!\left[{Y Z\over 8}\right.
  &+\cos(\beta_0 a_w)\left(1+\frac{Y Z}{8}\right)\nonumber\\
  &\left.+{j\over 2}\sin(\beta_0 a_w) \left(\frac{Z}{Z_0}+\frac{Y Z_0}{2}\right)\right].
\end{eqnarray}
Note that the dispersion equation for the Bloch waves is qualitatively
different from the dispersion equation for the surface waves on a
homogenized mushroom layer (for example, compare it with Eq. (20)
from~\cite{Yakovlev2009}). Nevertheless, we will soon see that the
location of the propagation band predicted by the Bloch model
coincides with the one predicted by the dispersion equation for the
TM-polarized surface waves, $k_t = k_0\sqrt{1 - (1/Y_s^{TM})^2}$,
\cite{Yakovlev2009}. Indeed, near the resonance of the HIS,
$1/Y_s^{TM}\rightarrow \pm i\infty$ and thus $k_t\gg k_0$, from which
one may conclude that the surface waves must appear near the HIS
resonance. However, in addition to the dispersion characteristics, the
Bloch wave approach allows us to find the required termination
impedance, as is shown next.

In Eq.~(\ref{dispersion_equation_kx}), $Z_0$ is the reference TL
impedance and $\beta_0=k_0\sqrt{\varepsilon_r}$ is the TL propagation
factor.  The propagation factor in the direction orthogonal to an MS
edge is $k_{\perp} = k_{x'}\sqrt{2}$. Thus, the Bloch impedance $Z_B$
for the respective surface modes reads~\cite{Maslovski2018}
\begin{equation}
\label{bloch_impedance}
Z_B = \pm \left(Z_0 \tan\frac{\beta_0\, a_w}{2}-\frac{j Z}{2}\right) \cot\frac{k_{\perp} a_w}{2\sqrt 2},
\end{equation}
where the sign must be chosen so that $\Re{(Z_B)}>0$.

\begin{figure}
	\centering
	\includegraphics[width=0.45\textwidth]{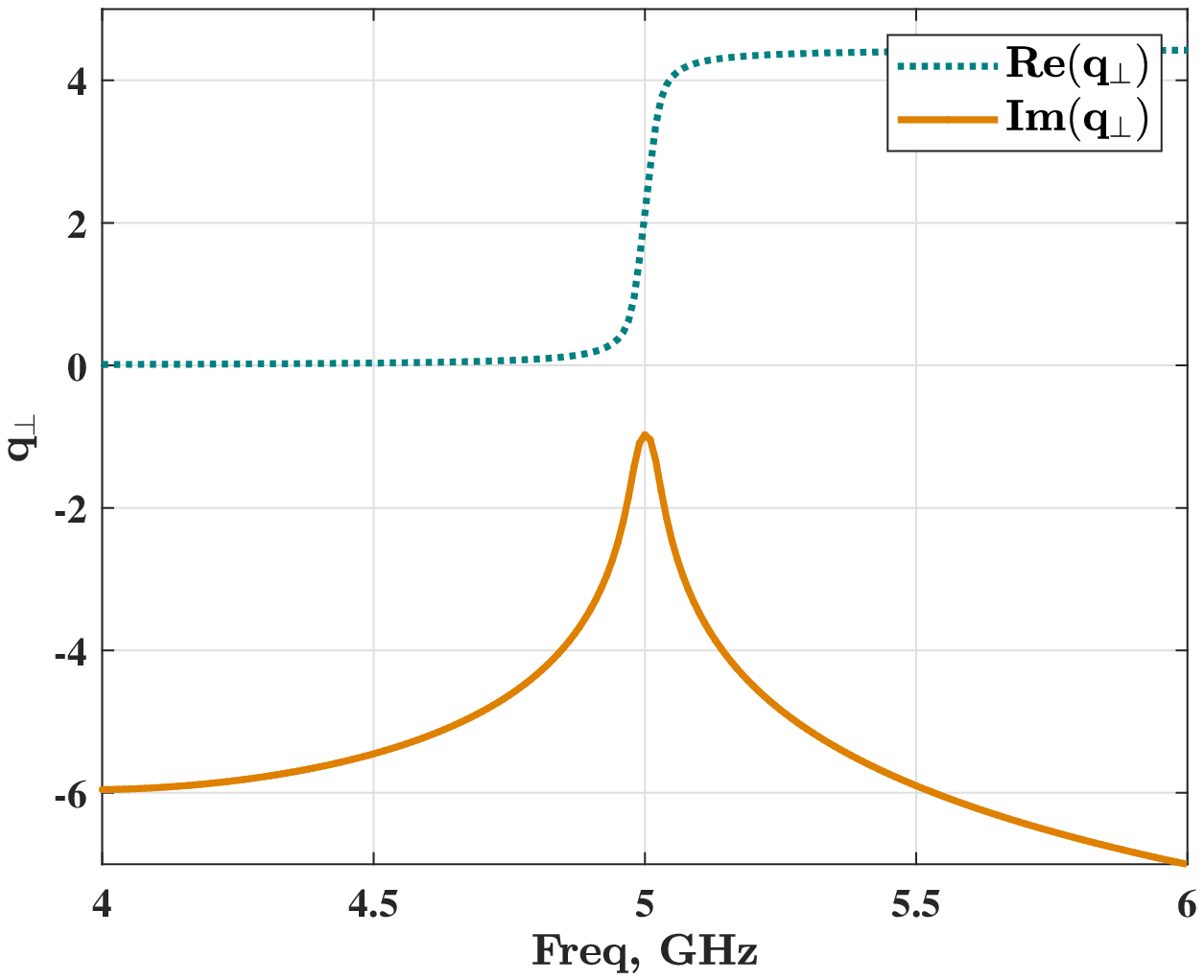}
	\includegraphics[width=0.45\textwidth]{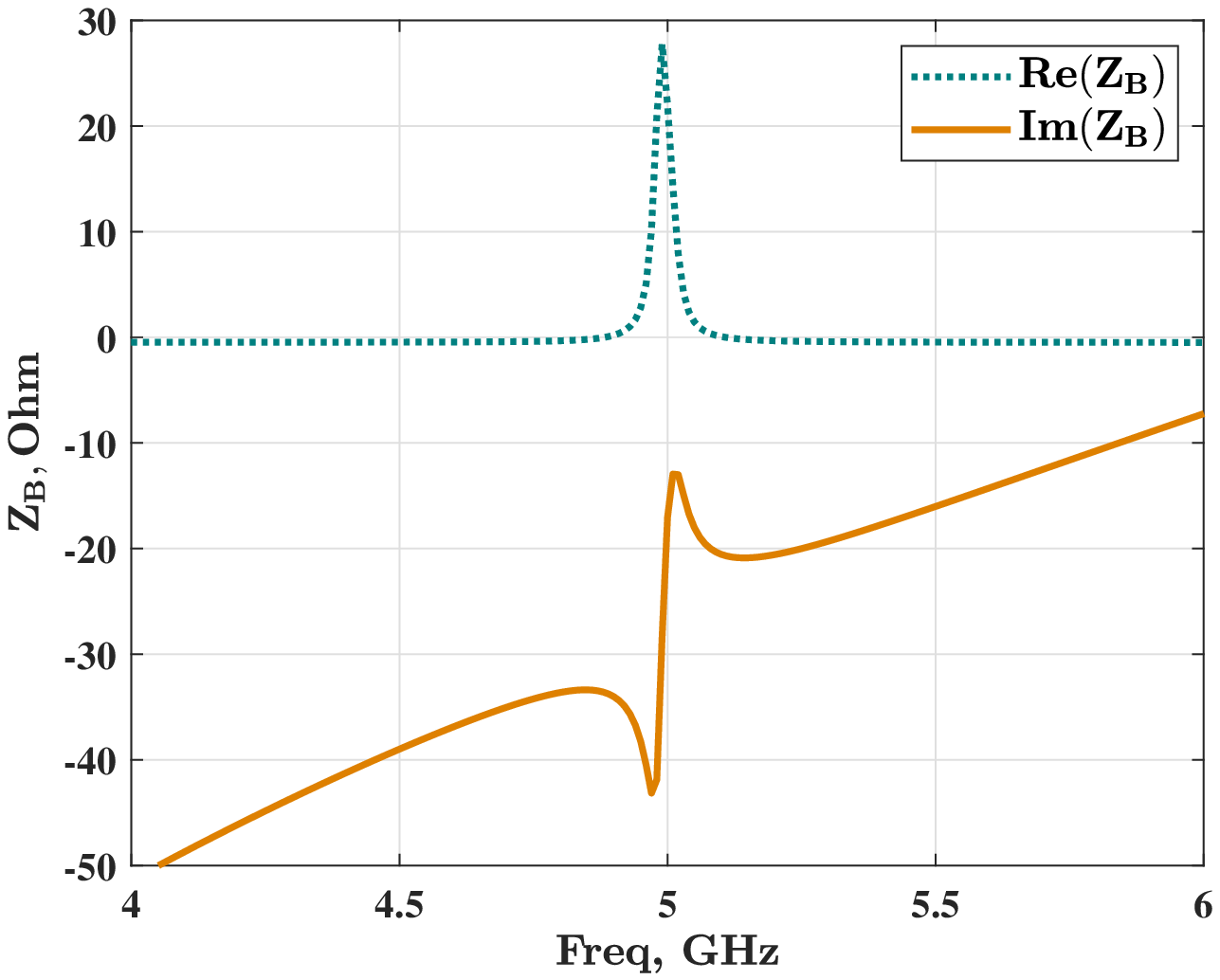}\\
	\caption{Panel (a): Dispersion of the Bloch waves, where
		$q_{\perp}=k_{\perp}a_{wm}$ is the normalized propagation
		factor. Panel (b):
		Bloch wave impedance $Z_B$ as a function of frequency.}
	\label{fig:qd}
\end{figure}

To estimate the deteriorating effect of the resonant Bloch modes
excited on the PMS, we have modeled finite MS samples with the unit
cell size and other parameters as in the previous calculations for the
infinite periodic MS. For the purpose of the Bloch impedance and
dispersion calculations, the reference impedance $Z_0$ was estimated
from the substrate parameters and the width of the top metalization at
the points where the varactors are connected to the patches:
$Z_0=71.6~\Omega$, by using the standard formulas for microstrip
lines. The varactor diode model used in the following studies was
characterized with the serial inductance $L_s = 0.2$~nH, the serial
resistance $R_s = 0.88$~Ohm, and the package capacitance
$C_p \approx 0.14$~pF. The varactor junction capacitance was set to
$C_j = 0.2$~pF in order to obtain the HIS resonance at around 5~GHz.

The obtained dispersion of the parasitic Bloch waves is plotted in
Figure~\ref{fig:qd}(a), and the Bloch impedance as a function of
frequency is shown in Figure~\ref{fig:qd}(b). It can be seen that our
qualitative analytical model predicts existence of the propagating
Bloch waves near 5 GHz, with the real part of the Bloch impedance
close to 20--30 Ohm in the middle of the band. Based on these results,
we conclude that in order to suppress the unwanted surface mode
resonances, one has to add terminating resistors. One possibility is
to add such resistors at the connections of the vias to the ground
plane at the edges of the MS. Because at such edge the $x'$- and
$y'$-directed currents of the rotated unit cells
(Figure~\ref{fig:TL_circ}) add up, the terminating resistance must be
set to $Z_B/2$. Because the impedance varies within the band, a
roughly average value of $Z_B/2 \approx 11$~Ohm was selected.

\begin{figure}
	\centering
	\subfigure[]{\label{fig:surfwave2_a}\includegraphics[width=0.45\textwidth]{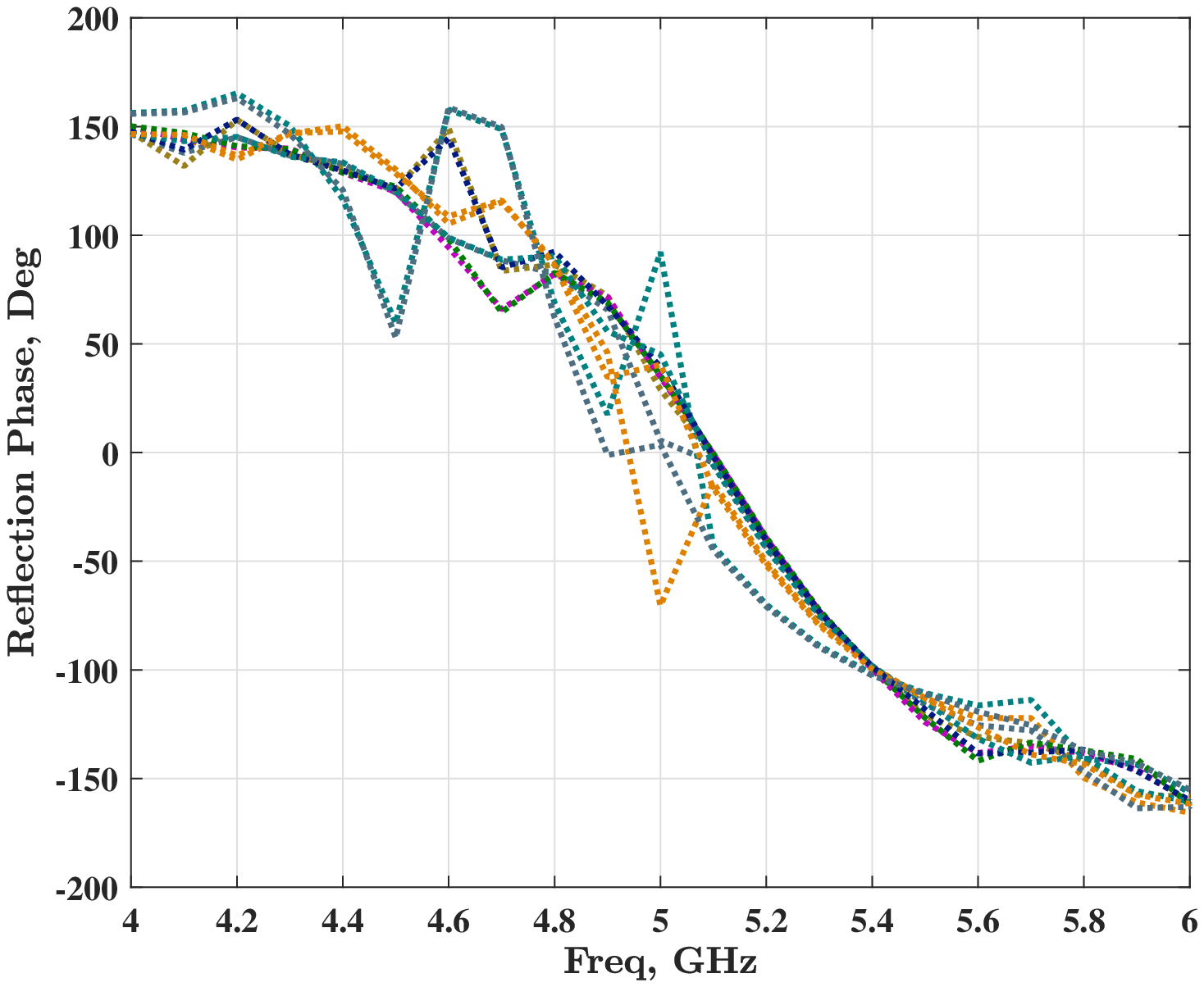}}
	\subfigure[]{\label{fig:surfwave2_b}\includegraphics[width=0.45\textwidth]{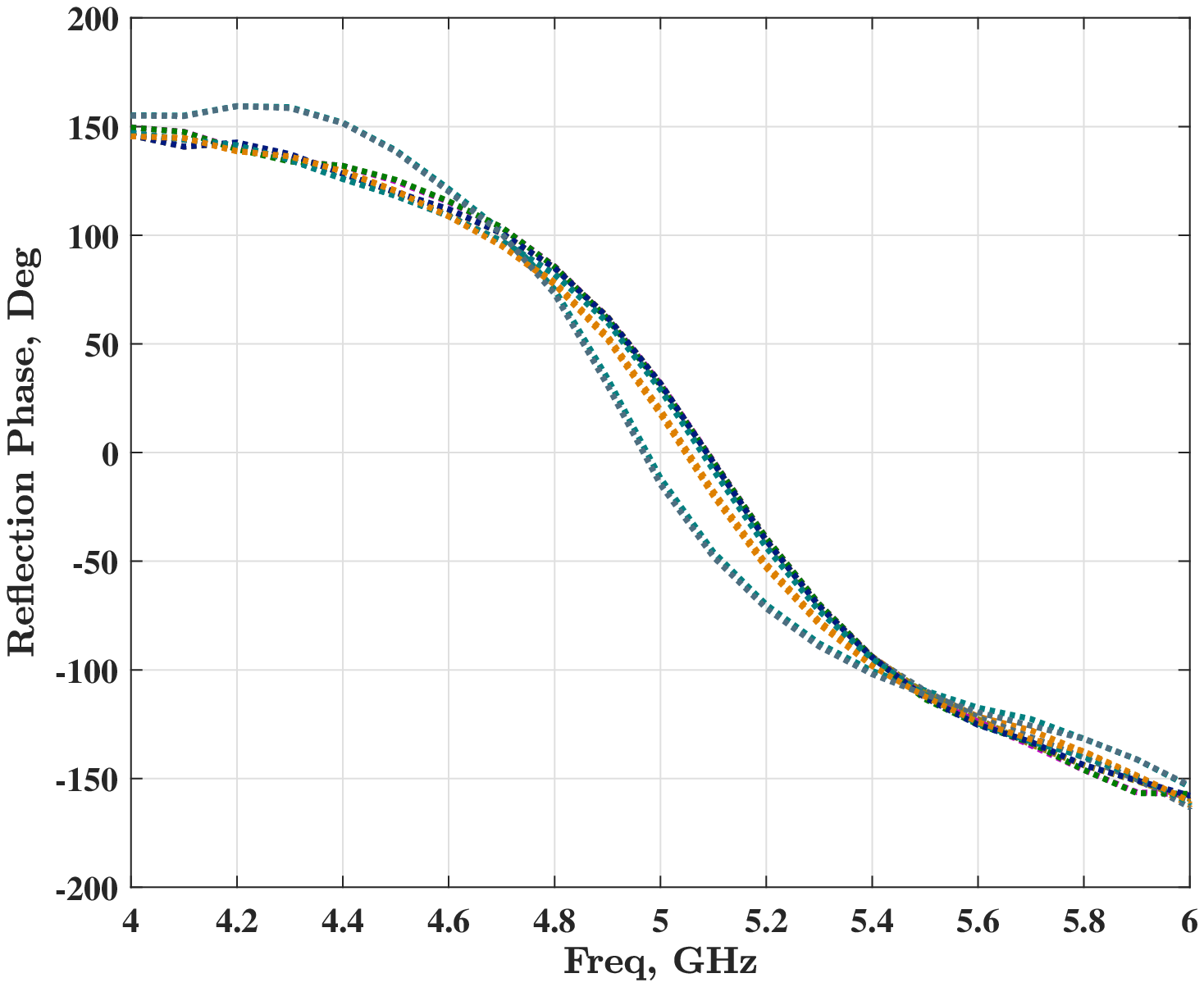}}
	\caption{Parasitic mode suppression in a finite-size MS formed
          by $3\times10$ unit cells (SIMULIA CST Studio Suite
          simulation; several curves are for the local reflection
          phases at different unit cells). Panel (a): The phase
          performance of the finite-size MS before adding the $Z_B/2$
          terminating resistors. Panel (b): The phase performance of
          the finite-size MS after adding the $Z_B/2$ terminating
          resistors.}
	\label{fig:surfwave2}
\end{figure}

To demonstrate effectiveness of the terminating resistors for
suppression of parasitic modes, we have numerically simulated several
finite-size MS samples in SIMULIA CST Studio Suite. Some
characteristic results are shown in Figure~\ref{fig:surfwave2} before
and after adding the termination resistors. From
Figure~\ref{fig:surfwave2_a} one can see that in the MS sample without
the terminating resistors the parasitic modes are excited in the
frequency range close to 5 GHz, which affects the phase distribution
over the MS. One may also notice the unwanted resonances at the
frequencies below 5 GHz, which are not immediately accounted by the
qualitative model. However, from Figure~\ref{fig:surfwave2_b} one can
see that all parasitic resonances disappear in the finite MS sample
with the terminating resistors. This confirms that the proposed method
of parasitic mode suppression with the terminating resistance obtained
from the Bloch wave model is viable and effective.

\section{Design and modeling of the controlling network with capacitive
  memory}
\label{control_circuit}

In this section we describe an extension to the controlling network
formed by the $x$ and $y$ bias lines that allows us to set up the
varactor bias voltages individually for each unit cell of the MS.
In order to achieve this functionality, we add a memory capacitor to
each unit cell, as is explained next.

\begin{figure}
	\centering
	\includegraphics[width=0.9\linewidth]{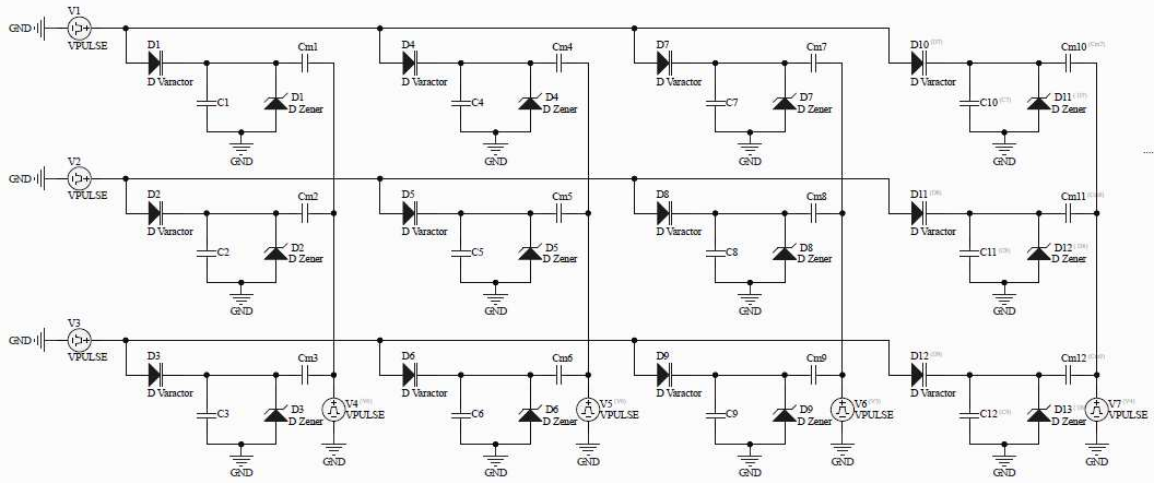}
	\caption{Controlling network with capacitive memory as modeled
          in ADS.}
	\label{fig:CL}
\end{figure}

The extended controlling network with capacitive memory is depicted in
Figure~\ref{fig:CL}. As compared to the basic bias network in which
the anodes of the varactors are simply connected to the $x$ bias lines
and the cathodes to the $y$ bias lines, in the extended network we
introduce additional memory capacitors $C_{{\rm m}, i}$,
$i = 1,2,\ldots$ (one per unit cell), which are inserted between the
$y$ controlling lines and the vias leading to the varactor's cathodes
[the corner vias in Figure~\ref{fig:spice_model}(a)]. Note that in the
equivalent network shown in Figure~\ref{fig:CL}, a single diode symbol
represents a pair of varactors from the original MS, because
these varactors are effectively connected in parallel with each other
at dc and share the same bias voltage. In the same equivalent network,
we also present the filter capacitors $C_i$, the purpose of which is
to prevent the high-frequency currents from flowing into the rest of
the bias network.

As can be seen from Figure~\ref{fig:CL}, at the intersection of the
control lines the varactor diode and the memory capacitor are
connected in such a way that when the varactor diode is forward-biased
the memory capacitor gets charged through the open diode (the
programming phase), and when the polarity is opposite the memory
capacitor keeps its charge and the voltage on this capacitor (in
addition to the standby voltage on the control lines) determines the
dynamic capacitance of the varactor and thus the resonant frequency of
the unit cell. The Zenner diodes are included in this network to be able to discharge
the memory capacitors by applying high voltage pulses with inverse
polarity to the $y$-lines. As we will show next, with such bias network topology we are able to
control each cell separately, which means that more freedom in
manipulating the local surface impedance and the reflection phase is
available.

The programming of this PMS with capacitive memory can be accomplished
by applying appropriate pulsed voltage to the structure, in which the
memory capacitors are repeatedly charged to a certain capacity, or
discharged through the Zenner diodes with erasing pulses. The
time-domain signal analysis of such programming and erasing processes
has been carried out by using Agilent ADS. In the simulations we
have considered the circuit from Figure~\ref{fig:CL} that includes a 3-by-4
network of the control lines that we name X1, X2, X3 (the horizontal
lines, counted from the top) and Y1, Y2, Y3, Y4 (the vertical lines,
counted from left to right).

\begin{figure}
	\centering
	\subfigure[]{\label{fig:gp_a}\includegraphics[width=0.45\textwidth]{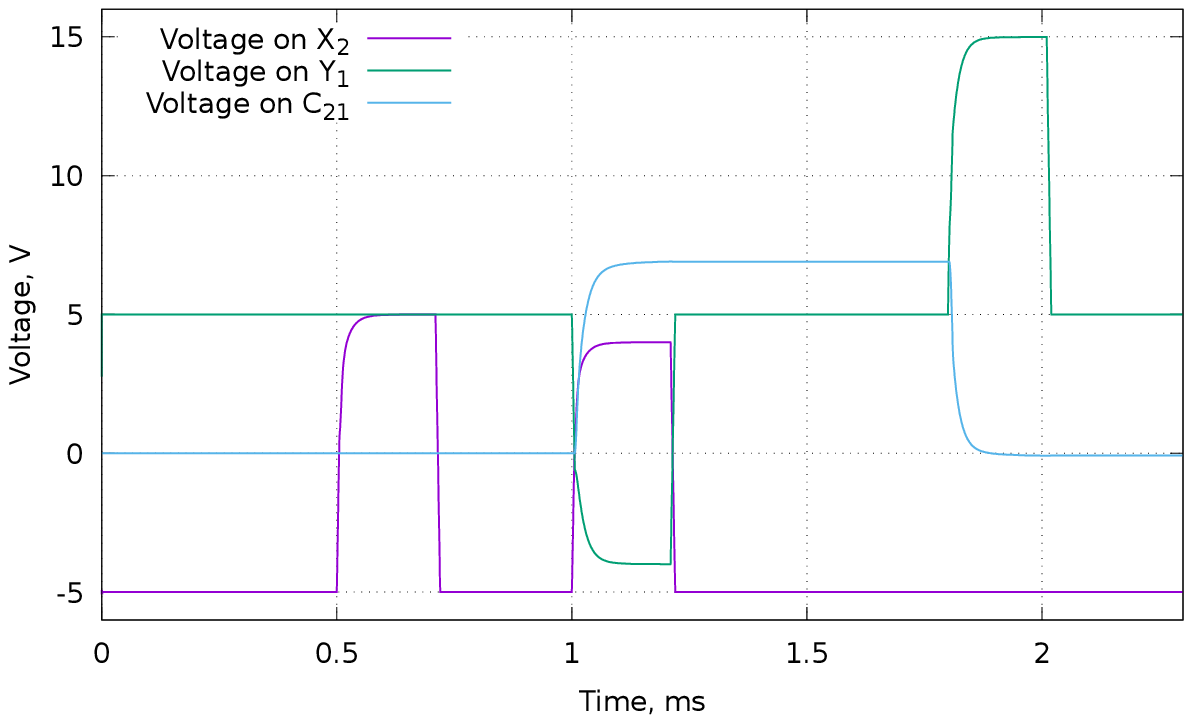}}
	\subfigure[]{\label{fig:gp_b}\includegraphics[width=0.45\textwidth]{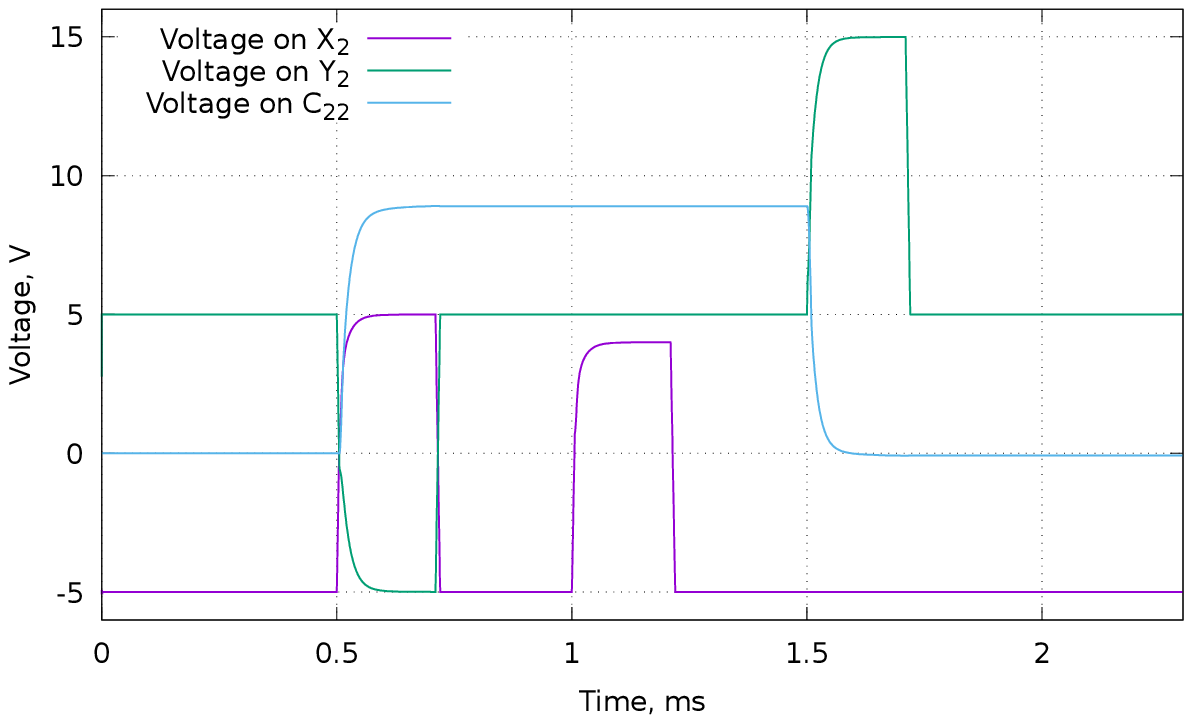}}
	\caption{Panel (a): Voltages on the X2 and Y1 lines and the
          memory capacitor $C_{21}$. Panel (b): Voltages on the X2 and
          Y2 lines and the memory capacitor $C_{22}$. The memory
          capacitor values are $C = 1$~uF, and the characteristic
          charging/discharging time is on the order of $RC$, where $R = 50$~Ohm are the
          current limiting resistors connected to all control lines
          (not shown in Figure~\ref{fig:CL}).}
	\label{fig:gp}
\end{figure}

In Figure~\ref{fig:gp_a} we plot the simulated voltages on the X2 and
Y1 lines and on the memory capacitor $C_{{\rm m}, 2} = C_{21}$
belonging to the $(2,1)$-th unit cell, and in Figure~\ref{fig:gp_b} we
show the voltages on the X2 and Y2 lines and on the capacitor
$C_{{\rm m},5}= C_{22}$ in the (2,2)-th unit cell. As one can see, the
coincident voltage pulses of opposite polarity on the lines X2 and Y2
(starting at $t = 0.5$~ms) charge the memory capacitor in the
$(2,2)$-th cell (programming phase). At the same time, the $(2,1)$-th
cell is not affected. Analogously, the coincident voltage pulses on
the lines X2 and Y1 (starting at $t = 1$~ms) program the $(2,1)$-th
unit cell, and the $(2,2)$-th cell is not affected. One may also note
that the programmed voltage on the memory capacitors can be adjusted
by varying the amplitudes of the voltage pulses on the X and Y
lines. That is why the programmed voltages on $C_{21}$ and $C_{22}$
are different. Finally, the erasing high voltage pulses starting at
$t = 1.5$~ms and $t = 1.8$~ms on the Y1 and Y2 lines discharge the
memory capacitors connected to the respective lines. Note that the
entire columns of unit cells connected to these lines can be erased
independently.

To conclude, we have confirmed that the proposed implementation of the
control network with memory capacitors enables independent programming
of unit cells of the PMS. Additionally, the proposed schematic reduces
component cost because here the varactors combine the functions of
diode switches and controllable capacitive loads.

\section{Conclusions}

In this study, a number of analytical and numerical methods to analyze
reconfigurable Sievenpiper mushroom-type PMS equipped with varactors
and memory capacitors have been developed. We have used these methods
to propose and investigate a PMS structure suitable for microwave
applications at the frequencies from $3.6$ to $6$ GHz (e.g., for the
future 5G+ systems). The unit cell design and the control network
topology proposed in this article enable independent programming of
all unit cells of the PMS.

The analytical model developed in Section~\ref{imped_model} has been
used to predict the surface impedance and the complex reflectivity of
the considered PMS for the TE and TM incidence cases with a high accuracy. In
Section~\ref{num_results}, the analytical results have been
validated with the full-wave numerical simulations in SIMULIA CST Studio
Suite for a wide range of frequencies and incidence
angles. Additionally, a simplified equivalent circuit model has been
developed and tested in Agilent ADS. The results of the analytical and
numerical models are in very good agreement and demonstrate potential
benefits of using the studied structures for the beamforming and
beam steering applications. In particular, the proposed chessboard-like
PMS has shown excellent stability of the resonant frequency with
respect to the incidence angle for both TE and TM polarizations.

In Section~\ref{Bloch_model}, we have developed a qualitative
analytical model for the parasitic resonant surface modes
on the finite-size mushroom-type PMS. With this model, we
have been able to determine the optimal value of the terminating
resistors that have to be added to the unit cells at the edges
of the structure in order to suppress unwanted resonances and
recover the expected beamforming behavior of the PMS.

In Section~\ref{control_circuit}, we have proposed and studied the
controlling network suitable for the considered PMS. In such network,
the controlling voltage in the form of short bipolar pulses is applied
to the unit cells through a network of disconnected $x$ and $y$
control lines and a set of memory capacitors. We have performed a
number of circuit simulations in order to determine optimal parameters
of the network and the amplitude and the duration of the programming
and erasing pulses applied to the coordinate lines. The obtained
results have confirmed the validity of our designs.

The theoretical models and practical concepts developed in this
article will find applications in the actively developing
areas of research and technology that deal with reconfigurable
metasurfaces for future telecommunication systems.

\section*{Acknowledgment}

This work has been funded by Funda\c{c}\~{a}o para a Ci\^{e}ncia e a
Tecnologia (FCT), Portugal, under the Carnegie Mellon Portugal Program
(project ref.~CMU/TIC/0080/2019).

\section*{References}


\end{document}